\documentclass[preprint]{aastex61}

\usepackage{amssymb, amsmath}

\DeclareSymbolFont{UPM}{U}{eur}{m}{n}
\DeclareMathSymbol{\umu}{0}{UPM}{"16}
\let\oldumu=\umu
\renewcommand\umu{\ifmmode\oldumu\else\math{\oldumu}\fi}
\newcommand\micro{\umu}

\newcommand\ttt[1]{10\sp{#1}}
\newcommand\tttt[1]{\times10\sp{#1}}

\usepackage{threeparttable}
\usepackage{booktabs}
\usepackage{chngcntr}

\shorttitle{Constraining the Dust Opacity Law}
\shortauthors{Webb {\em et al.}}

\begin{document}

\title{Constraining the Dust Opacity Law in Three Small and Isolated Molecular Clouds}

%% AUTHOR/INSTITUTIONS FOR AASTEX6.1:
\author{K.~Webb}
\affiliation{Department of Physics \& Astronomy, 3800 Finnerty Road, University of Victoria, Victoria, BC, V8P 5C2, Canada}

\author{J.~Di Francesco}
\affiliation{National Research Council of Canada, Herzberg Institute of Astrophysics, 5071 West Saanich Road, Victoria, BC, V9E 2E7, Canada}

\author{S.~Sadavoy}
\affiliation{Harvard-Smithsonian Center for Astrophysics, 60 Garden Street, Cambridge, MA 02138, USA}

\author{K.~Thanjavur}
\affiliation{Department of Physics \& Astronomy, 3800 Finnerty Road, University of Victoria, Victoria, BC, V8P 5C2, Canada}

\author{R.~Launhardt}
\affiliation{Max-Planck-Institut f{\"u}r Astronomy, K{\"o}nigstuhl 17, D-69117, Heidelberg, Germany}

\author{Y.~Shirley}
\affiliation{Steward Observatory, 933 North Cherry Avenue, Tucson, AZ 85721, USA}

\author{A.~Stutz}
\affiliation{Departmento de Astronom{\`i}a, Facultad Ciencias F{\'i}sicas y Matem{\'a}ticas, Universidad de Concepci{\'o}n, Av. Esteban Iturra s/n Barro Universitario, Casilla 160-C, Concepci{\'o}n, Chile\\}

\author{J.~Abreu Vicente}
\affiliation{Max-Planck-Institut f{\"u}r Astronomy, K{\"o}nigstuhl 17, D-69117, Heidelberg, Germany}

\author{J.~Kainulainen}
\affiliation{Max-Planck-Institut f{\"u}r Astronomy, K{\"o}nigstuhl 17, D-69117, Heidelberg, Germany}

%% AUTHOR/INSTITUTIONS FOR EMULATE APJ:
% \author{Patricio~E.~Cubillos\altaffilmark{1,2},
% Joseph~Harrington\altaffilmark{1},
% and
% Third~Author\altaffilmark{1}
% }
% \affil{\sp{1} Planetary Sciences Group, Department of
%               Physics, University of Central Florida, Orlando, FL 32816-2385\\
%        \sp{2} Space Research Institute, Austrian Academy of Sciences,
%               Schmiedlstrasse 6, A-8042, Graz, Austria}

\email{kawebb@uvic.ca}

% %% Extra info for aastex:
% \received{Yesterday}
% \revised{Today}
\accepted{September 27, 2017}
% \published{Tomorrow}
% \submitjournal{AASJournal}

\begin{abstract}
Density profiles of isolated cores derived from thermal dust continuum emission rely on models of dust properties, such as mass opacity, which are poorly constrained. With complementary measures from near-infrared extinction maps, we can assess the reliability of commonly-used dust models. In this work, we compare \emph{Herschel}-derived maps of the optical depth with equivalent maps derived from CFHT WIRCAM near-infrared observations for three isolated cores: CB\,68, L\,429, and L\,1552. We assess the dust opacities provided from four models: OH1a, OH5a, Orm1, and Orm4. Although the consistency of the models differs between the three sources, the results suggest that the optical properties of dust in the envelopes of the cores are best described by either silicate and bare graphite grains (\textit{e.g.}, Orm1) or carbonaceous grains with some coagulation and either thin or no ice mantles (\textit{e.g.}, OH5a). None of the models, however, individually produced the most consistent optical depth maps for every source. The results suggest that either the dust in the cores is not well described by any one dust property model, the application of the dust models cannot be extended beyond the very center of the cores, or more complex SED fitting functions are necessary.
\end{abstract}

\section{Introduction}\label{sec:introduction}

The initial density and temperature of dense regions (`cores') of molecular clouds (MC) which collapse to form stars heavily influence the process of star formation. At the typical cold temperatures and high densities of these dusty star-forming systems, the primary constituent, hydrogen, cannot be directly observed, and the pre-stellar conditions are therefore measured through alternative tracers. The three principal methods for probing these conditions are: 1) dust emission, 2) dust extinction, and 3) emission from molecules. In this paper, we concentrate on the consistency of information obtained from dust.

The MCs in the solar neighborhood have temperatures in the range of 8-50 K, resulting in thermal dust emission that is most easily detected at far-infrared (FIR) and submillimeter (submm) wavelengths \citep{lada07}. The line-of-sight (LoS) dust column density may be determined by fitting a modified black-body to the spectral energy distribution (SED) of the thermal continuum emission. This technique, however, requires knowledge of the dust properties which are difficult to constrain due to a degeneracy with the grain size distribution, and changing physical conditions in the MCs \citep[\textit{e.g.},][]{reach95, henning96, agladze96, draine03, boudet05, coupeaud11}. In addition, the SED fits to thermal dust emission are a product of the dust column density and dust temperature. modeling thermal dust emission is complicated by the fact that the emission changes with varying temperatures along the LoS \citep{shetty09a, shetty09b}. Full three-dimensional radiative transfer models can be used to characterize the dust better, but such models can be very complex and require significant assumptions about the cloud geometry. 

The extinction of background starlight by dust absorption in the near-infrared (NIR) accurately traces the dust distribution \citep{nice}. The colors of stars are reddened when observed through dust which is directly related to the hydrogen content via a reddening law \citep[\textit{e.g.},][]{rieke85, bohlin78}. Together with an observationally-determined constant gas-to-dust mass ratio \citep[\textit{e.g.},][]{lilley55, bohlin78, sodroski97}, extinction mapping provides a direct link between the dust and total cloud mass. Dust extinction is also independent of the LoS dust temperature (to first order) and the uncertainties from unknown dust properties are generally smaller than those obtained from the emission method. On the other hand, extinction mapping relies on the need for a statistically significant surface density of background stars and an a priori knowledge of the intrinsic stellar colors. It is therefore typically not well suited to map the centers of dense cores where extinction (and column density) is high (A$_V > 30$) and the number of background stars is low.

Furthermore, dust grains are expected to change within the dense core and through its evolution. Dust evolves mainly through grain growth, where gas particles accrete onto grain surfaces to form icy mantles, or small grains coagulate to form larger grains \citep{ossenkopf94, stepnik03, paradis05, ormel11}. The optical properties of the dust grains, such as the mass opacity, change with grain composition, structure, and size. At present, `standard' dust models \citep[such as ][]{ossenkopf94, weingartner01, ormel11} are typically adopted to describe dust properties. 

To gain better leverage of the dust properties of clouds and cores, we combine independent measurements of column density (optical depth) from dust emission and absorption using different dust models. We study three isolated cores, CB\,68, L\,429, and L\,1552 using archival FIR \emph{Herschel}\footnote{\emph{Herschel} is an ESA space observatory with science instruments provided by European-led Principal Investigator consortia and with important participation from NASA.} observations of dust emission and recently obtained high-resolution NIR observations measuring dust extinction. Although our primary focus is on assessing the dust opacity model OH5a, which was used in the ``The Earliest Phases of Star Formation'' \citep[EPoS,][]{stutz10} analysis, we will also consider other common dust models.

We describe the NIR observations and summarize the FIR \emph{Herschel} observations in Section \ref{sec:observations}. In Section \ref{sec:analysis}, we summarize (as it is well-documented elsewhere) the methods used to produce optical depth maps from the \emph{Herschel} data, and describe the methods used to produce optical depth maps from the extinction maps. The comparisons of the optical depth ratios to the dust model opacity ratios is discussed in Section \ref{sec:discussion}. Finally, in Section \ref{sec:conclusions}, we outline our conclusions.

\begin{table*}
\centering
\begin{threeparttable}
\caption{Properties of the cores in this survey. Each are isolated, nearby, and well characterized by previous studies.}
\label{tab:targets}
\begin{tabular}{cccccc}
\toprule
  Source	& R.A., Dec, (J2000)	& Region & Distance (pc) 	& Size (\arcmin) & Evolution class 	\\
 \midrule
  CB\,68 (L 146)	& 16:57:16, -16:09:18	& Ophiuchus	& 120 $\pm$ 20 (1)	& 24$\times$52 (2)		& protostar (1)	\\
  L\,429			& 18:17:05, -08:13:32	& Aquila		& 225 $\pm$ 55 (3)	& 18$\times$12 (2)		& starless (3)	\\
  L\,1552			& 05:17:39, +26:05:05	& Taurus		& 140 $\pm$ 10 (4)	& 19.8$\times$28.2 (2)	& starless (5)  	\\ 
\bottomrule
\end{tabular}
\begin{tablenotes}
      \footnotesize
      \item Notes: L, \citet{lynds62}; CB, \citet{clemens88}
      \item References: \textbf{1.} \citet{launhardt97}, \textbf{2.} \citet{dobashi05}, \textbf{3.} \citet{stutz09_l429}, \textbf{4.} \citet{kenyon94}, \textbf{5.} \citet{stutz09}
      \end{tablenotes}
\end{threeparttable}
\end{table*}

\begin{figure*}[ht]
  \begin{center}
  \includegraphics[height=2.2in]{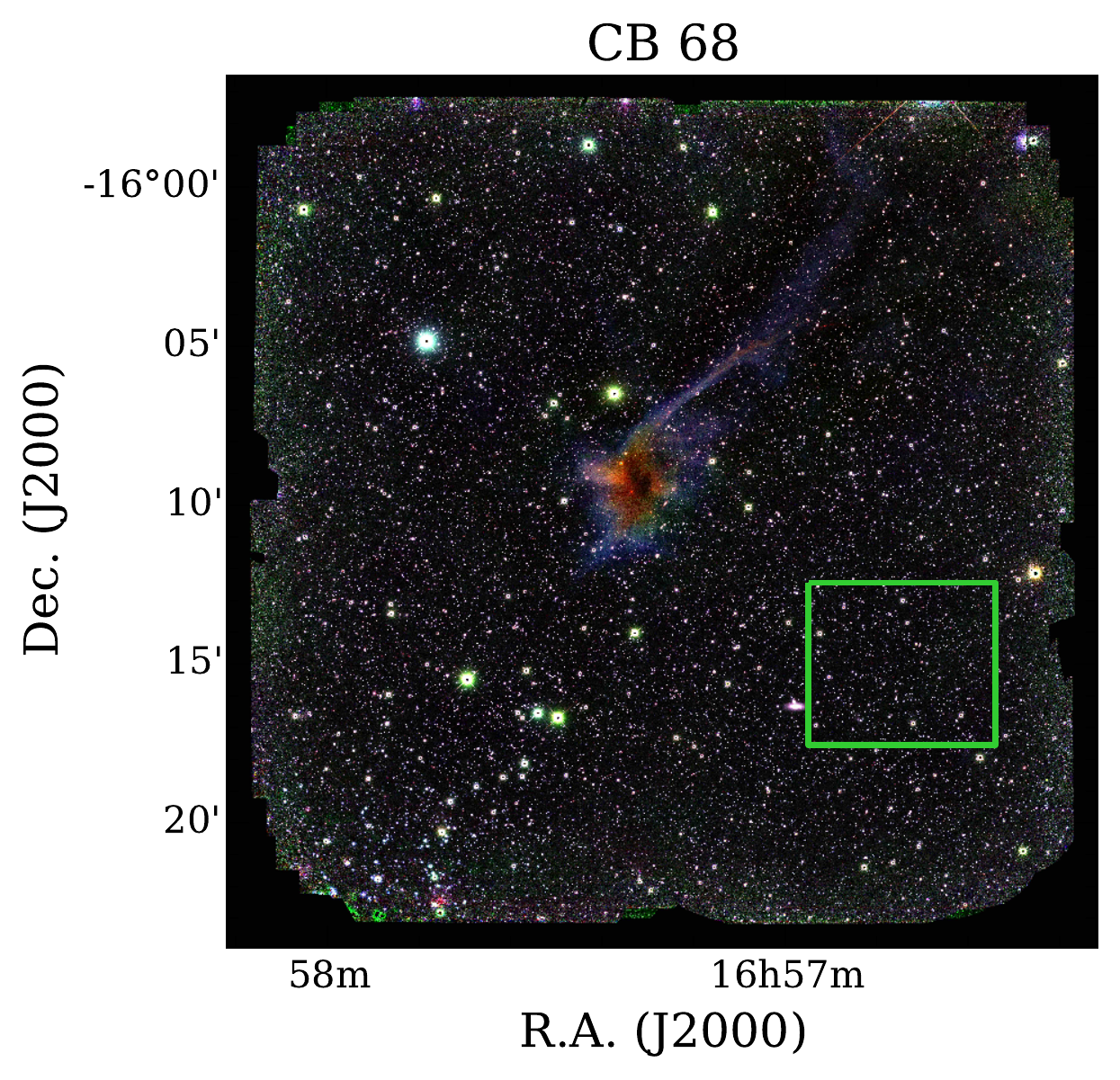}
  \includegraphics[height=2.2in]{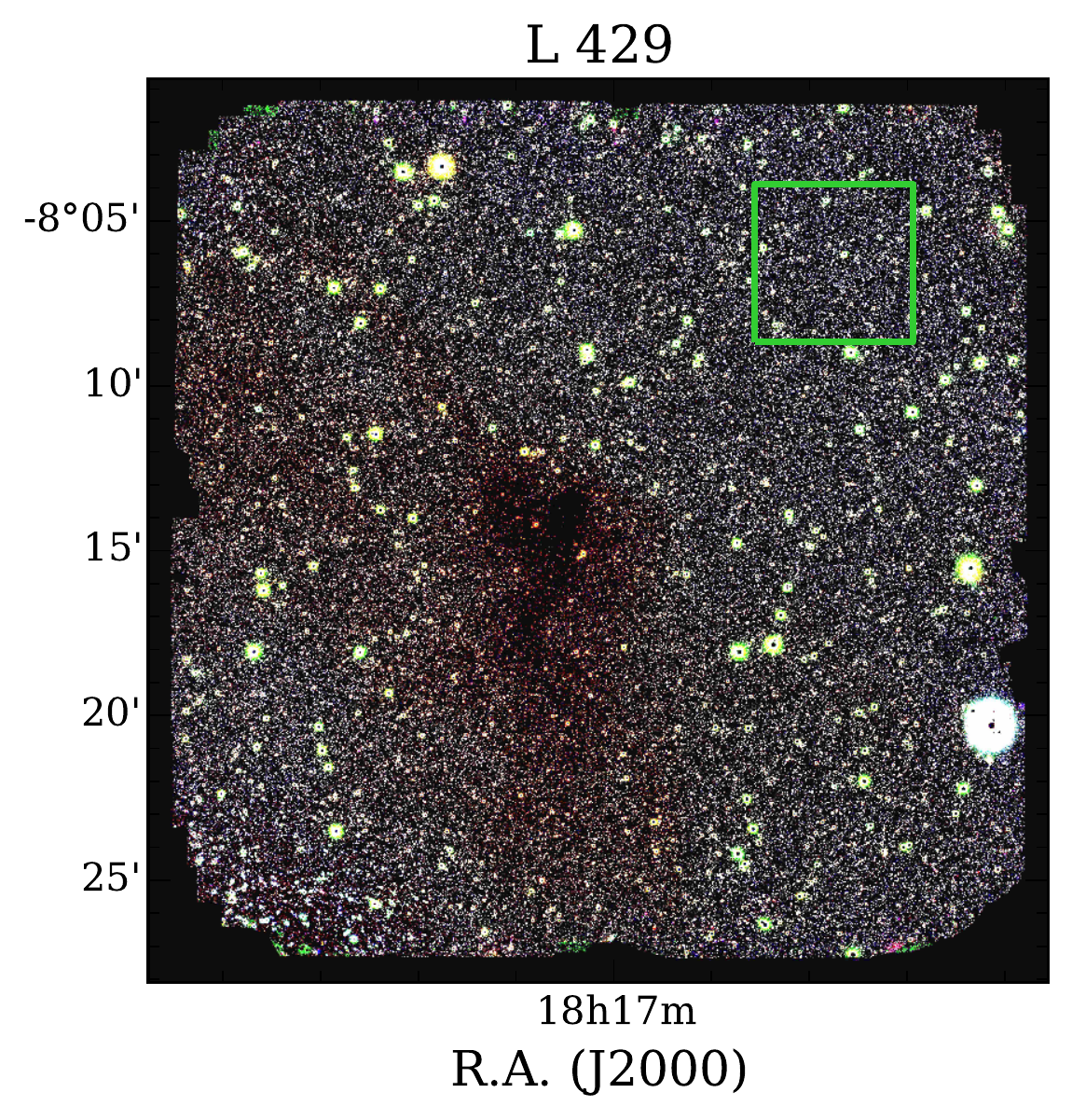}
  \includegraphics[height=2.2in]{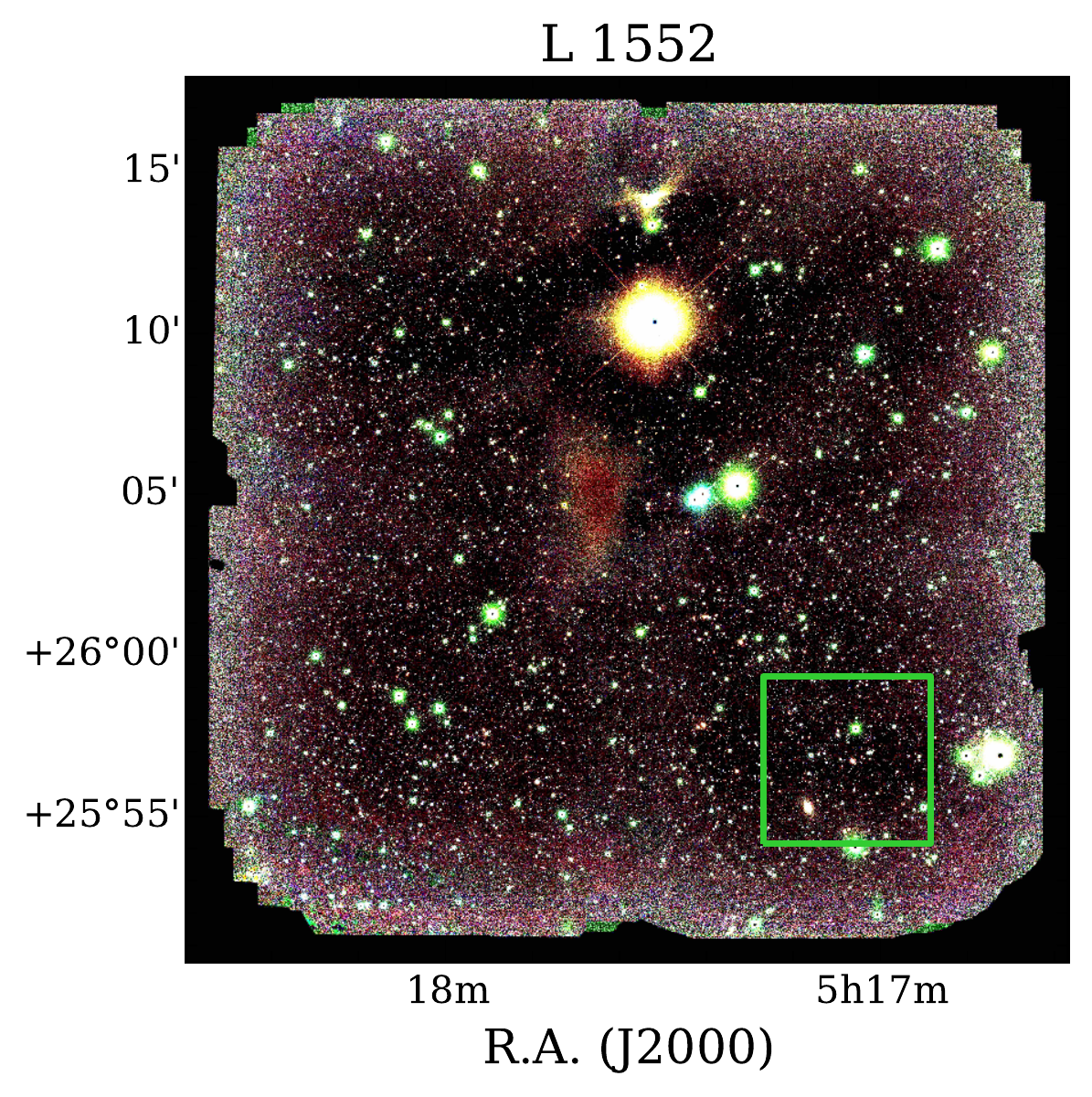}
  \end{center}
  \caption{RGB (JHKs) false-color images of the cores. Regions with minimal extinction used to calibrate the intrinsic color scatter in the background stars are outlined in green.   } \label{fig:rgb}
\end{figure*}

%%%%%%%%%%%%%%%%%%%%%%%%%%%%%%%%%%%%%%%%%%%%%%%%%%%%%%%%%%
%%%%%%%%%%%%%%%%%%%%%%%%%%%%%%%%%%%%%%%%%%%%%%%%%%%%%%%%%%

\section{OBSERVATIONS}
\label{sec:observations}

We selected three isolated, nearby, and previously well-characterized cores: CB\,68, L\,429, and L\,1552. The position, associated region, and size of each core is provided in Table\,\ref{tab:targets}. The targets were selected from regions with low FIR background confusion noise, and with sufficiently high surface densities of background stars to derive deep NIR extinction maps. Initial results of modeling indicated that the structures of these cores are relatively simple and resolved enough to allow for reliable dust property analysis, therefore reducing the error on the measured extinction. L\,429 and L\,1552 are both starless, whereas CB\,68 has an embedded low-luminosity protostar \citep{launhardt10} with a simply structured envelope that is sufficiently extended to allow for the comparison with the dust emission-derived map. With this sample, we probe different environments which are easily modeled.

\subsection{\emph{Herschel} observations} \label{sec:herschel}

FIR continuum observations were obtained from the \emph{Herschel} Space Observatory \citep{pilbratt10} which covers the wavelength range 100-500 $\micro$m using its two bolometers: the Photodetector Array Camera and Spectrograph \citep[PACS;][]{poglitsch10}, observing at 100 $\micro$m and 160 $\micro$m (HPBW=7.1$\arcsec$ and 11.2$\arcsec$, respectively), and the Spectral and Photometric Imaging Receiver \citep[SPIRE;][]{griffin10}, observing at 250 $\micro$m, 350 $\micro$m, and 500 $\micro$m (HPBW=18.2$\arcsec$, 24.9$\arcsec$, and 36.3$\arcsec$, respectively).

The Herschel observations were taken from the Herschel Science Archive (HSA). We used the Level 2.5 data products for PACS and Level 2 products for SPIRE.  All data were reduced with HIPE using version 14 of the the HSA pipeline. The PACS data were imaged using the Scanamorphos routine \citep{roussel13}, whereas the SPIRE data were imaged using the Naive Mapper in HIPE with the optional correction for extended emission. More details about the HSA data reduction can be found in the Data Reduction Guides for both instruments\footnote{http://herschel.esac.esa.int/hcss-doc-13.0/load/spire\_drg/html/spire\_drg.html}$^\mathrm{,}$\footnote{http://herschel.esac.esa.int/twiki/pub/Public/PacsCalibrationWeb/pdrg\_phot\_Hipe14.pdf}.

Absolute sky fluxes were not measured by \emph{Herschel}, and as a result, \emph{Planck} and \textit{IRAS} data have been often used to determine zero-point flux offsets for the \emph{Herschel} bands \citep[\textit{e.g.}, ][]{lombardi14}. The Level\,2 SPIRE maps from the \emph{Herschel} archive apply a similar correction using \emph{Planck} data at 857\,GHz and 545\,GHz (see the SPIRE Handbook for more details). PACS data from the archive, however, have no such correction. Therefore, we estimate the zero-point flux offset in the PACS data using the all-sky \emph{Planck} SED models in a similar manner as \citet{lombardi14}. The \emph{Planck} SED models provide maps of dust temperature and optical depth (at 353\,GHz) at 5\,$\arcmin$ resolution and dust emissivity, $\beta$, at 30\,$\arcmin$ resolution. With these SED parameters, we constructed modified blackbody functions for each \emph{Planck} pixel and then determined the corresponding flux that would have been measured by the \emph{Herschel} bands. Unlike \citet{lombardi14}, we cannot simply convolve the \emph{Herschel} data to a similar resolution as the \emph{Planck} models (\textit{e.g.}, 5\,$\arcmin$) and compare fluxes pixel by pixel, because the \emph{Herschel} maps of the globules are relatively small. For example, the PACS maps are generally less than 10\,$\arcmin$ in size. Instead, we compared azimuthal averages of the \emph{Planck}-determined emission maps at each of the 160-500\,$\micro$m \emph{Herschel} bands to azimuthal averages of the observed \emph{Herschel} data, where it is expected that the two radial profiles are equal at large radii.

The SPIRE maps were indeed in good agreement with those expected from \emph{Planck}, illustrating the success of their zero-point corrections. The radial emission profiles at 250\,$\micro$m, 350\,$\micro$m, and 500\,$\micro$m show good agreement with expectations from \emph{Planck} at large radii (e.g., $>$ 300$\arcsec$). The PACS 160\,$\micro$m profiles, however, had substantial deficits relative to the expected 160\,$\micro$m profiles  at the same large radii. Since we expect PACS data to also match \emph{Planck} data at the same radii as SPIRE data, we apply a correction on the 160\,$\micro$m data to bring the two profiles into agreement. The PACS 160\,$\micro$m offsets were $\sim$45\,MJy\,sr$^{-1}$ for CB\,68, $\sim$160\,MJy\,sr$^{-1}$ for L\,429, and $\sim$73\,MJy\,sr$^{-1}$ for L\,1552. Thus, our radial profile correction technique provides a good, first-order estimate of the zero-point offsets at 160\,$\micro$m. A full study of this method will be given in Sadavoy et al. (in preparation). 

The SPIRE data covered a larger area than those from PACS for all three cores. As a result, each core has a central region where all \emph{Herschel} wavelengths were observed, and an outer region where only SPIRE wavelengths were observed. The optical depths in the SPIRE-only regions have much larger uncertainties due to the missing constraints on the fitted SED on the Wien side of the spectrum. For this reason, only the regions with both SPIRE and PACS coverage were considered in the following comparison. 

Column densities from optically thin thermal dust emission will trace all the emission along the line of sight. In contrast, column densities from dust extinction will primarily trace the foreground component of dust associated with the nearby globules themselves. Thus, it is necessary to subtract a ``background'' column density level for the thermal dust emission so that these data better match the dust extinction observations. To be conservative, we use the median background emission associated with radii $>$\,300\,$\arcsec$ from the central core. This background measurement is likely overestimated as it will include some material from the core as well. Extensive modeling to determine how to best correct the thermal dust emission, however, is beyond the scope of this paper. A discussion of the results of the analysis without applying this correction is provided in Appendix\,\ref{app:background}.

% here1

\subsection{CFHT WIRCam} \label{sec:wircam}

Observations of dust extinction were obtained from the Canada-France-Hawaii Telescope (CFHT) Wide-field Infrared Camera \citep[WIRCam;][]{puget04}\footnote{www.cfht.hawaii.edu/Instruments/Imaging/WIRCam/} in the J, H, and K (1.25\,$\micro$m, 1.63\,$\micro$m, and 2.15\,$\micro$ m, respectively) NIR wavebands. CFHT WIRCam has a 20$\arcmin$$\times$20$\arcmin$ field-of-view (FOV) well suited to observe each of the cores with a single pointing, while allowing for simultaneous observations of surrounding sky fields with minimal extinction. The instrument is mounted on the 3.6\,m CFHT, and has a focal plane consisting of four 2048$\times$2048 pixel detectors, with a sampling of 0.3$\arcsec$ pix$^{-1}$. The cores were imaged between Sept.-Oct. 2014 under program 14BC37, and May 2015 under program 15AC18 (P.I.: J. Di Francesco). A summary of the integration times, sizes, and depths of the final observations is listed in Table\,\ref{tab:obs_nir}. The exposures were dithered to correct for the CCD chip gaps and bad pixels, and combined with the CFHT preprocessing pipeline IDL Interpreter of WIRCam Images (`I'iwi). A background image was produced from the stacked exposures, then subtracted from each exposure. Post-processing was performed at the \textit{Terapix}\footnote{terapix.iap.fr} astronomical data reduction center at the Institut d'Astrophysique in Paris, where the images were astrometrically aligned and coadded. Scattered light from the dust in the densest regions, called `coreshine' \citep{steinacker10}, was observed in the cores of CB\,68 and L\,1552 \citep{stutz09, lefevre14} and care was taken to not subtract it from the images. Using the coreshine observations to constrain the dust models are beyond the scope of this paper, however. Figure\,\ref{fig:rgb} shows RGB-color images (for J, H, and Ks) of the final core maps where the effect of extinction is seen in the reddening toward the center of each core.

Aperture photometry was performed on the final images with the program Source Extractor (SExtractor, v2.19.5; \citealp{bertin96}). Detections at a threshold of 7\,$\sigma$ (minimum area threshold of 5\,pix, and Gaussian filter) were matched to objects in the Two-Micron All-Sky Survey\footnote{www.ipac.caltech.edu/2mass/} \citep[2MASS;][]{2mass} catalogue to determine the zero-point-magnitude. The detection threshold was then lowered to 3.5\,$\sigma$ for the same detection parameters, where anomalous detections were removed in the object cross-match in the catalogues of the three filters. Stellar objects were selected from the linear distribution of half-light radii as a function of magnitude, and the colors were calculated for objects in common between the three NIR bands. 

The depth of the survey was assessed through the insertion of artificial stars where the limiting magnitude was taken as the magnitude at which 50\% of a sample of artificial stars was recovered by the detection pipeline (see Table\,\ref{tab:obs_nir}). 

The depth of the observations suffered from a known issue with the Direct Imaging Exposure Time Calculation (DIET)\footnote{http://www.cfht.hawaii.edu/Instruments/Imaging/WIRCam/dietWIRCam.html}, and were less deep than predicted for the observing conditions and exposure times requested in the CFHT proposal. Sufficient numbers of stars were detected, however, to achieve a resolution in the NIR extinction maps equal to the \emph{Herschel} optical depth maps. The significance of not having the intended depth is discussed further in Section \ref{sec:magdiff}.

Some extended emission was observed in the SPIRE maps in the selected `unreddened' regions of L\,429 and L\,1552 (see Section\,\ref{sec:extinction_maps}). Therefore, to remove any bias introduced by an unclean background, we applied the same background subtraction procedure as used for the emission maps (discussed in Section\,\ref{sec:herschel}). We note that the background extinction subtraction was insignificant in comparison with the background emission subtraction.\\

\begin{table*}
\centering
\begin{threeparttable}
\caption{Summary of the WIRCam observations.}
\label{tab:obs_nir}
\begin{tabular}{cccccccc}
\toprule
  Source & 
  \multicolumn{3}{c}{Exposure time (s)} & 
  Size ( $^\prime \, \times \, ^\prime$ ) &
  \multicolumn{3}{c}{Limiting magnitude (mag)}\\
   &  J  & H & K &   & J & H & K \\
  \midrule
  CB\,68     & 2451	& 1425	& 2616	& 24.2$\times$24.2	& 20.4	& 19.1	& 19.4	\\
  L\,429      & 1710	& 1890	& 1968	& 30.8$\times$30.8	& 19.1	& 18.1	& 17.6	\\
  L\,1552    & 1710	& 1890	& 1272	& 24.2$\times$24.2	& 20.7	& 19.8	& 19.3	 \\
\bottomrule
\end{tabular}
\begin{tablenotes}
      \footnotesize
      \item Notes: Exposure times for standard WIRCam J, H, and Ks exposure times of 57\,s, 15\,s, and 24\,s, respectively. Comparing the total on-source times with the completeness values we have estimated, it appears that the depth in L\,429 is at least a magnitude shallower than the other two fields even though it was observed for an equivalent amount of time. As the magnitude is zeropoint corrected against 2MASS stars, the reason for the shallower depth could be perhaps that the seeing was significantly worse during the L\,429 observations. Although the image depths do not necessarily correspond to the relative exposure times, they are consistent with depths determined from an overturn in the number counts.
    \end{tablenotes}
\end{threeparttable}
\end{table*}

%%%%%%%%%%%%%%%%%%%%%%%%%%%%%%%%%%%%%%%%%%%%%%%%%%%%%%%%%%
%%%%%%%%%%%%%%%%%%%%%%%%%%%%%%%%%%%%%%%%%%%%%%%%%%%%%%%%%%

\section{ANALYSIS}
\label{sec:analysis}

\subsection{Optical depth maps from emission SED fitting} \label{sec:emission_maps}

The emission maps were convolved to the beam of the SPIRE 500\,$\micro$m map with the program MIRIAD\footnote{bima.astro.umd.edu/miriad/userguide\_US.pdf}, then regridded to 14\,$\arcsec$\,pix$^{-1}$ using Nyquist sampling. The average 1\,$\sigma$ uncertainties in the zero-point-corrected maps amounted to: 19\% for CB\,68, 14\% in L\,429, and 16\% in L\,1552.

$S_\nu$, the flux density as a function of frequency, $\nu$, in the data cubes, can then be described as:
\begin{equation} \label{eqn:sed}
S_\nu = \Omega \left( 1 - e^{-\tau_\nu} \right) B_\nu (T_d)
\end{equation}
\noindent where $ B_\nu (T_d)$ is the Planck function which is dependent on the dust temperature $T_d$ and frequency $\nu$,  $\Omega$ is the beam solid-angle, and $\tau_\nu$ is the optical depth through the cloud where 
\begin{equation} \label{eqn:tau_fir}
\tau_\nu = N_\mathrm{H} \, m_\mathrm{H} \frac{M_\mathrm{d}}{M_\mathrm{H}} \kappa_{\mathrm{d}, \, \nu}
\end{equation}
\noindent The optical depth is a function of the total hydrogen column density $N_H = 2 N(H_2) + N(H)$, the proton mass $m_H$, the dust-to-hydrogen mass ratio $M_d$\,/\,$M_H$ (assumed to be 1/110 following for example, \citeauthor{sodroski97} \citeyear{sodroski97}), and the dust mass absorption coefficient (here called the opacity), $\kappa_{d, \, \nu}$. Given that the \emph{Herschel} wavelength coverage is limited such that both the dust emissivity and temperature cannot be robustly measured simultaneously, we assumed a fixed dust grain model with a prescribed $\kappa_{d, \, \nu}$. Section \ref{sec:dust_models} describes the four dust models tested in this paper. 

A $\chi^2$ minimization technique was then used to fit an SED (describing a single temperature modified blackbody) to each pixel in the calibrated dust emission maps, where T$_d$ and $\tau_\nu$ were free parameters. We remind the reader that four different dust models were used, and we note that the slope of $\tau_\nu$ varies with the assumed dust model. The individual fluxes in the SED were weighted by $\sigma^{-2}$. Here the $\sigma$ is calculated as the quadrature sum of i) the uncertainty in the \emph{Planck}-derived offset and ii) the mean rms noise. Calibration errors and color correction errors were also accounted for using a Monte Carlo analysis.

In using 250\,$\micro$m as the fiducial wavelength for our comparison of the optical depths derived from the FIR, the column density can be determined via
\begin{equation}
N_\mathrm{H} = \left(6.58\tttt{25}\,\mathrm{cm}^{-2}\right) \tau_{250} / \left( \kappa_{250} / 1 \,\mathrm{cm}^2 \mathrm{g}^{-1}\right)
\end{equation}

\subsubsection{Dust opacity models} \label{sec:dust_models}

Dust models determine opacities for particular environments. It is therefore important to select an appropriate dust model for the conditions of an MC core. In fact, studies of coreshine \citep[\textit{e.g.},][]{steinacker14, lefevre14, lefevre16} suggest that a single dust opacity model is not appropriate. Rather, there is differential coagulation and ice mantle growth between the cold dense core interiors and the thin warm outer envelopes. Therefore, there can be large systematic uncertainties in the fits of the SEDs related to the dust opacity and its spectral dependence. This problem highlights the motivation of this paper: to test the accuracy of selected dust models used to measure column densities via comparison with supplementary NIR observations. 

We tested four commonly used dust opacity models which cover a wide range of dust properties.  We refer to these models as OH1a and OH5a from \citet[][OH]{ossenkopf94}\footnote{\label{ft:models}These dust models are slightly different than those tabulated in \citet{ossenkopf94} that assume coagulation at a density $\ttt{6}$ cm$^{-3}$, which is larger than what is typically observed in these cores, $\ttt{5}$ cm$^{-3}$. The models are available at vizier.cfa.harvard.edu/viz-bin/VizieR?-source=J/A+A/291/943, and include the albedo.}
 and Orm1 and Orm4 from \citet[][Orm]{ormel11}\footnote{Models made available at w.astro.berkeley.edu/\textasciitilde ormel/software.html differ slightly from the published values.}. 

The first model, OH1a, describes interstellar medium (ISM)-like dust. The grains are of silicate-based carbonaceous composition with no ice mantles and no coagulation. Extinction optical depths were determined from the albedos, $a$, provided by \citet{weingartner01} for R$_\mathrm{V}$=3.1,
\begin{equation}
\kappa_\mathrm{ext} = \kappa_\mathrm{abs} / (1-a)
\end{equation}
\noindent In contrast, OH5a grains were coagulated for $\ttt{5}$ years with gas densities of $\ttt{5}$ cm$^{-3}$. This model was used in \citet{launhardt13} for CB\,68. Again, the extinction opacities were used.

The Orm1 and Orm4 models describe mixed silicate and graphite grains of fraction 2:1 and ice mantle thickness coagulated for $\ttt{5}$ years. Orm1 has bare silicate and graphite grains, Orm4 has ice-silicate and graphite grains mixed at the level of individual aggregates. Both of these models describe the extinction optical depths, and are therefore consistent with the altered OH models.

\begin{figure}
  \begin{center}
  \includegraphics[width=0.5\linewidth]{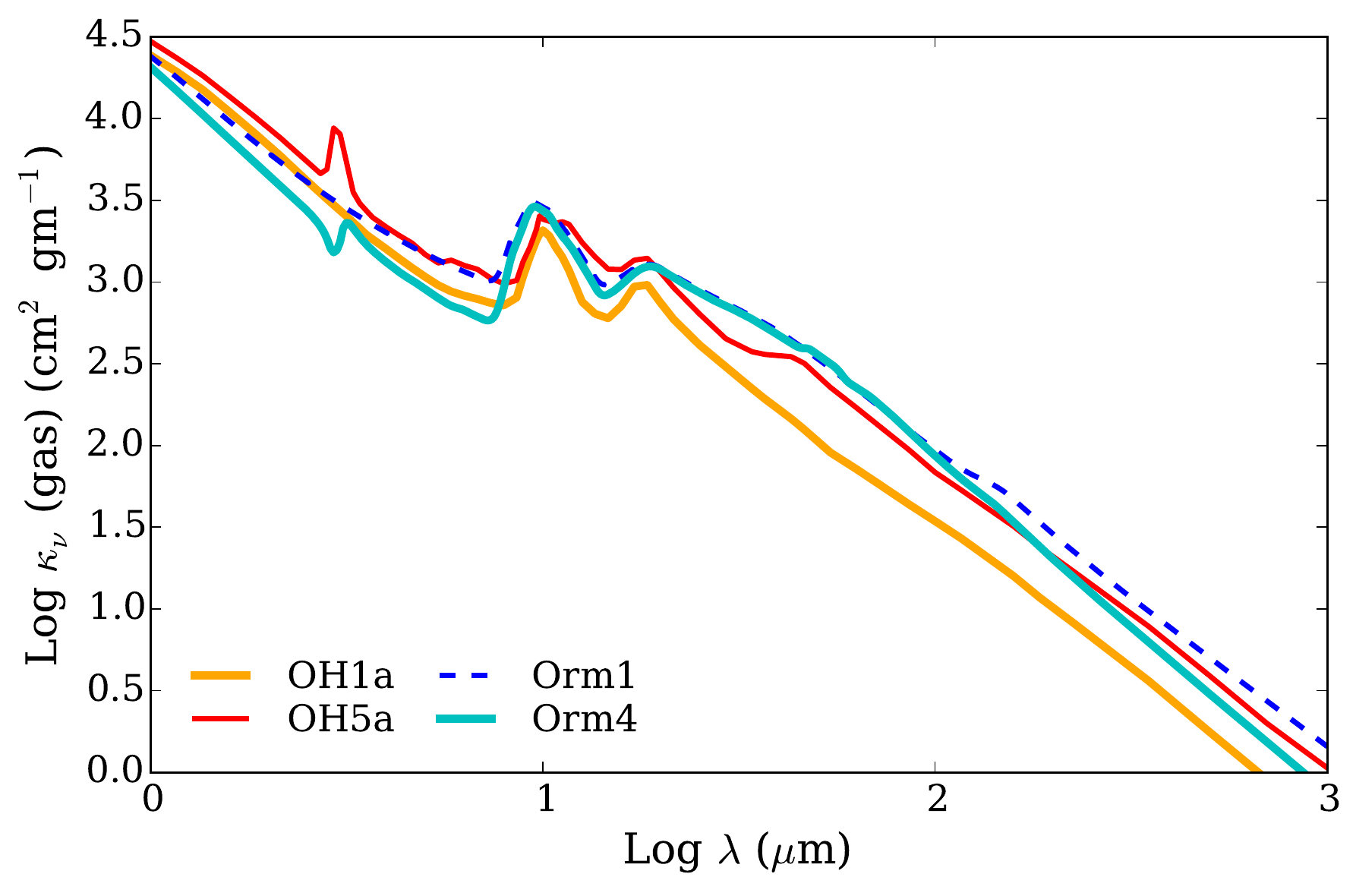}
  \end{center}
  \caption{Absorption coefficients of dust opacity for dust models in Table\,\ref{tab:consistency}. Scattering is only negligible when grains are much smaller than the wavelength (\textit{i.e.}, FIR and mm).} \label{fig:kappa_bounds}
\end{figure}

Figure\,\ref{fig:kappa_bounds} shows the predicted dust opacities for each of these models from NIR to submm wavelengths and Table \ref{tab:kappas} lists the values of the dust opacities at the NIR 1.25\,$\micro$m (J) wavelength and the FIR 250\,$\micro$m wavelength. Although the models have similar slopes in the 160-500\,$\micro$m range, there are differences in the ratios of the NIR and FIR opacities. The ratio $\kappa_\mathrm{J}/\kappa_{250}$ is 2.54$\tttt{3}$ for OH1a, 1.48$\tttt{3}$ for OH5a, 8.44$\tttt{2}$ for Orm1, and 1.09$\tttt{3}$ for Orm4. Four additional models, two each from OH and Orm, are discussed in Appendix \ref{app:other_models}.

One caveat to using the above models is that the densities describe the central cores of the sources, whereas our analysis is limited to the envelope where densities are $\ttt{3}$-$\ttt{4}$\,cm$^{-3}$. At lower densities dust will take longer to coagulate such that the dust sizes may be overestimated, which affects the determined optical properties. None of the models described in \citet{ossenkopf94} or \citet{ormel11}, however, include models at these lower densities.  \\

\begin{table}
\centering
%\begin{threeparttable}
\caption{Dust optical depth values for the models considered. Values are interpolated }
\label{tab:kappas}
\begin{tabular}{ccc}
\toprule
  Model	& $\kappa_\mathrm{J}$ [$\tttt{4}$ cm$^2$g$^{-1}$]	& $\kappa_{250}$ [cm$^2$g$^{-1}$] 	\\
 \midrule
OH1a	& 1.722	& 6.77	 \\
OH5a	& 2.098	& 14.16	 \\
Orm1	& 1.545	& 18.29	 \\
Orm4	& 1.728	& 15.84	 \\
\bottomrule
\end{tabular}
%\end{threeparttable}
\end{table}

\subsection{Optical depth maps from NIR extinctions} \label{sec:extinction_maps}

\begin{figure*}[ht]
  \begin{center}
  \includegraphics[width=\linewidth]{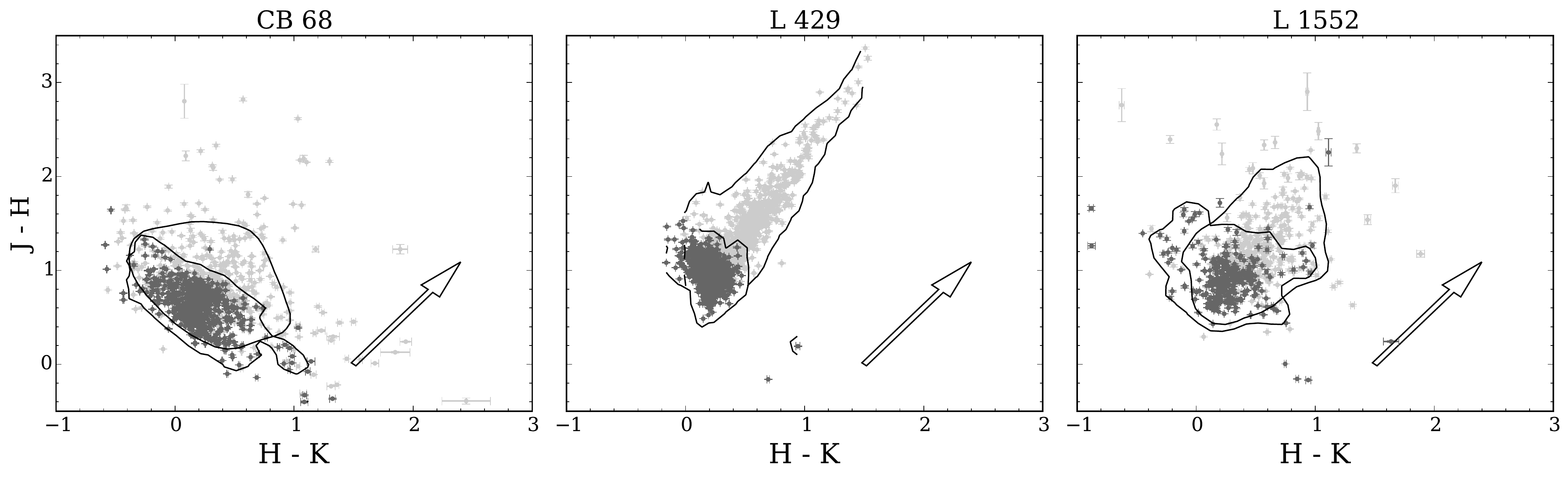}
  \end{center}
  \caption{NIR color-color distributions of stellar objects for CB\,68, L\,429, and L\,1552. Contours are shown at the 5 $\sigma$ level, where $\sigma$ is the standard deviation of the (J-K) -- (H-K) colors, divided by the square-root of the number of measurements. These statistics give an indication of the density and spread of the points. The effect of extinction is to shift the distribution along the reddening vector, shown as an arrow. The stars in the control regions (shown in Figure\,\ref{fig:rgb}) are shown in a darker hue, where it is apparent the reddening is negligible. Stars seen through the center of the core are shown in light grey.} \label{fig:colorcolor}
\end{figure*}

The extinction of background stars was used to derive independently the optical depth from NIR observations. The NIR color excess technique from \citet{nice} was used to combine the color information for the NIR bands with a standard reddening law to obtain a robust measure of the extinction in the cores. The extinction is proportional to the color excess which describes the observed flux S$_\lambda$ in one band $\lambda$ and a second $\lambda^\prime$, for a reddened (observed) and unreddened (intrinsic) field,
\begin{equation}
A_\lambda = E\left(S_\lambda - S_{\lambda^\prime} \right) / R_\lambda = \left( S_\lambda - S_{\lambda^\prime} \right)^{ obs} - \left( S_\lambda - S_{\lambda^\prime} \right)^{ intr}
\end{equation}
\noindent The color variations in a population of reddened and un-reddened stars in each core map are shown in Figure\,\ref{fig:colorcolor}. The effect of extinction is to shift the color-color distribution along the reddening vector, indicated by the arrow.

The proportionality constant R$_\lambda$ is dependent on the environment along the LoS. Empirical studies \citep[see the review by][]{mathis90} have shown that at visual wavelengths R$_\mathrm{V}$\,=\,3.1 for the ISM and as high as 5.5 in dense regions such as MCs. Recent studies such as \citet{schlafly16}, however, suggest not all MCs have high R$_\mathrm{V}$ values. A higher R$_\mathrm{V}$ is generally associated with dust grain growth via accretion and coagulation \citep[\textit{e.g.,}][]{weingartner01, flaherty07}. In the NIR, however, reddening laws \citep[such as][]{rieke85, cardelli89, mathis90} are independent of the environment such that A$_\mathrm{J}$$\sim$constant.

For the reddening law of \citet{cardelli89}, for R$_\mathrm{V}$=3.1,
\begin{equation*}
\tau_K = 0.600 \, \tau_H = 0.405 \, \tau_J
\end{equation*}
\noindent where $\tau_K = 0.114 \, \tau_V$. There is only a 15.6\% difference in $\tau_K$/$\tau_V$ if R$_\mathrm{V}$=5.5 is instead used.

The NIR color excess method of \citet[][NICEST]{nicest} was used to map the extinctions, where the technique has been tailored from those of \citet{nice} and \citet{nicer} to correct for the inhomogeneous distribution of background stars and its effect on smoothing small-scale structure in the optical depths. The application of smoothing parameters related to the distribution of stars corrects for the bias towards regions of low extinction by a factor of the sample size. 

Table\,\ref{tab:nicer_params} lists the color variations for stars in these control regions. Intrinsic color variations were determined from regions with minimal reddening. Although extended emission was observed in these control regions in the SPIRE maps of L\,429 and L\,1552, the effect of the bias is completely dominated by the approximation of the background subtracted from both the emission and extinction maps. Refer to Appendix\,\ref{app:background} for a discussion of the significance of this bias when no background subtraction is performed.

The final extinction map (A$_\mathrm{J}$) was determined from an unbiased and minimum variance maximum-likelihood combination of the individual extinctions, binned to match the resolution of the FIR emission-derived optical depth maps. The sensitivities of the extinction maps are directly related to the number of background stars in each pixel and the uncertainty of the measured magnitude. Measurements in locations with few background stars, due to high extinctions or an inhomogeneous background stellar distribution, are reported with higher relative error. A minimum of 10 stars per pixel was applied to ensure robust measurements. The smoothing parameter used to produce the extinction maps was selected to match the sampling (14$\arcsec$ pix$^{-1}$)\footnote{The pixel scale was chosen to optimize both the number of pixels and the majority of pixels meet the minimum 10 star criteria.} of the binned \emph{Herschel} optical depth maps, with the consequence of having several locations with an insufficient number of background stars to calculate the extinction with confidence. Hence, there are pixels in the final maps without a measure of the extinction.

The optical depth maps were then determined from the simple relation, $\tau_{\lambda} = A_\lambda / 1.086. $\\

\begin{table}
\centering
%\begin{threeparttable}
\caption{Parameters used for the calculation of the extinction with NICEST. Intrinsic J-H, and H-K scatter of the control (minimal extinction) region stars, maximum J-band extinction, and median of the error in the J-band extinction.}
\label{tab:nicer_params}
\begin{tabular}{ccccc}
\toprule
  Source & 
  $ \left( J - H \right)^{intr}$ & 
  $ \left( H - K \right)^{intr}$ & 
  max A$_\mathrm{J}$ &
  med $\sigma$A$_\mathrm{J}$ \\
  & & & [mag] & [mag] \\
 \midrule
  CB\,68	& 0.58$\pm$0.19	& 0.11$\pm$0.18	& 4.44	& 0.093	\\
  L\,429	& 0.91$\pm$0.14	& 0.25$\pm$0.09	& 16.68	& 0.034	\\
  L\,1552	& 0.77$\pm$0.17	& 0.22$\pm$0.17	& 3.95	& 0.180	\\
\bottomrule
\end{tabular}
%\end{threeparttable}
\end{table}

%%%%%%%%%%%%%%%%%%%%%%%%%%%%%%%%%%%%%%%%%%%%%%%%%%%%%%%%%%
%%%%%%%%%%%%%%%%%%%%%%%%%%%%%%%%%%%%%%%%%%%%%%%%%%%%%%%%%%

\section{DISCUSSION \& RESULTS}
\label{sec:discussion}

\subsection{Assumption of constant LoS temperature} \label{sec:temperature_maps}

SED fits to thermal dust emission are sensitive to the estimates of dust temperature, as discussed in Section \ref{sec:emission_maps}. In Equation \ref{eqn:sed}, the modified blackbody function is fit with a single temperature, which is representative of the average temperature along the LoS. These fits do not take into account any temperature gradients. For example, the cores will be heated by the interstellar radiation field. CB 68 in particular is protostellar, so it has a central heating source. Radiative transfer models of filaments and spherical clouds with central sources by \citet{ysard12} show that temperatures were overestimated by single-temperature SED fits compared to the true dust temperatures. Several other studies \citep[\textit{e.g.},][]{pagani15, lippok16, steinacker16} also point out issues in modeling dense cores with single-temperature blackbody functions. For diffuse media, the column density can be underestimated by up to a factor of 1.7 in \emph{Herschel} bands and 2.2 in \emph{Planck} bands. Such a bias would be most significant where the gradient is strongest and least significant where the temperatures are more uniform. Moreover, it will vary from source to source (\textit{e.g.}, the factor will vary with the magnitude of the gradient and the distribution of material). 

In the context of this paper, an overestimation of dust temperature up to a factor of $\sim$1.7 diminishes the measured $\tau_{250}$ based on analysis by \citet{ysard12}. The optical depth will primarily be underestimated toward the center of the core. This material, however, is not traced in the extinction maps and was excluded from our analysis. The following comparison of optical depths are therefore reasonable, despite the simplifying assumption of a constant LoS temperature. 

\begin{figure*}[ht]
  \begin{center}
  \includegraphics[width=\linewidth]{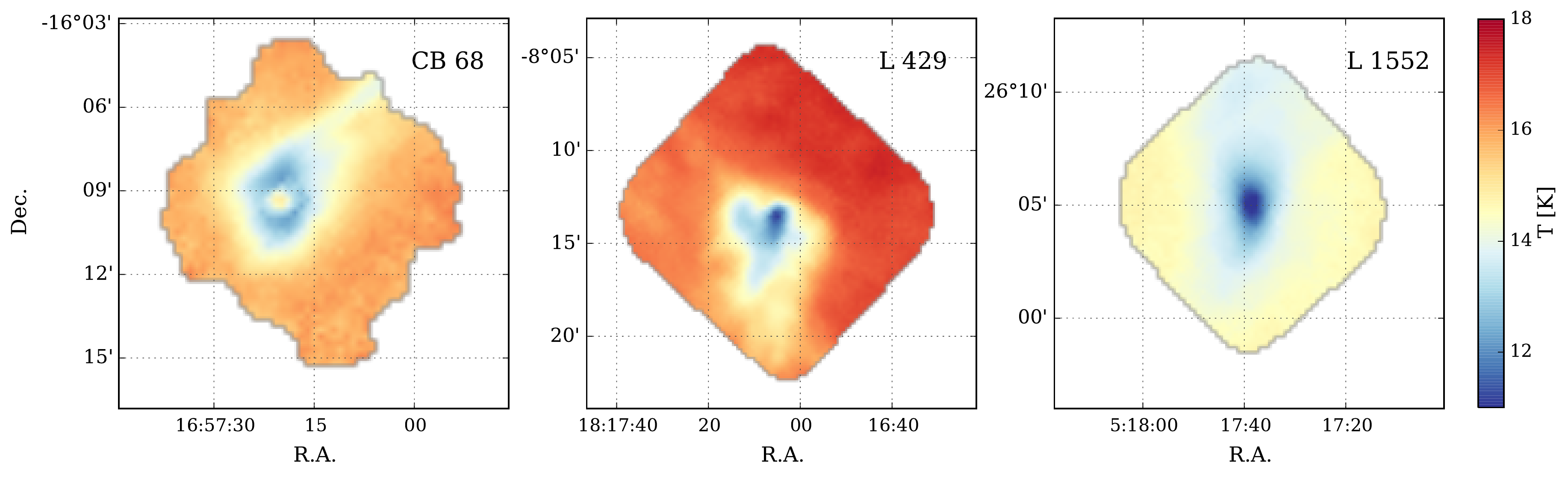}
  \end{center}
  \caption{FIR thermal dust continuum emission SED fit temperature maps for CB\,68, L\,429, and L\,1552 derived for OH5a. } \label{fig:tmaps}
\end{figure*}

Figure\,\ref{fig:tmaps} shows the dust temperature maps derived from the SED fitting procedure for the three cores. L\,429 has the steepest temperature gradient from its outermost edges to its center. By comparison, CB\,68 has a relatively small gradient, even though it is the only protostellar core in our sample. 

\subsection{Visual extinction sensitivities} \label{sec:Avs}

\begin{figure}[ht]
  \begin{center}
  \includegraphics[width=0.5\linewidth]{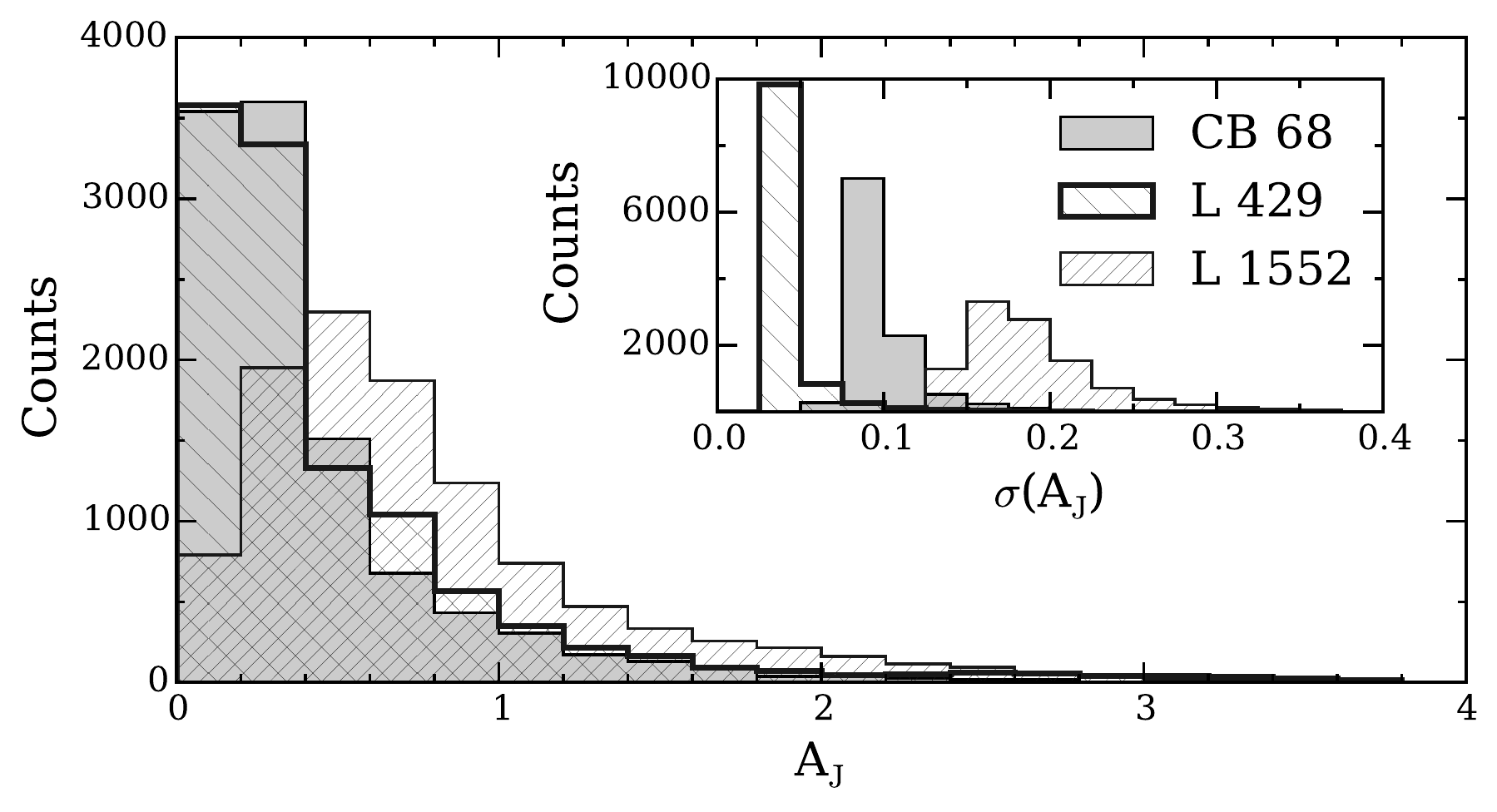}
  \caption{Histogram of visual extinction values measured in each core with bin size 0.2\,mag. The associated uncertainties of the extinctions are shown in the inset, with bin size 0.1\,mag.} \label{fig:Av_hist}
  \end{center}
\end{figure}

Figure\,\ref{fig:Av_hist} shows the distributions of extinction and associated uncertainties in each core. Table\,\ref{tab:nicer_params} lists the maximum A$_\mathrm{J}$ and median rms sensitivities of the cores. The map rms sensitivities range from $\sigma$A$_\mathrm{J}$$\sim$0.064--1.030\,mag in CB\,68, 0.027--0.592\,mag in L\,429, and 0.111--0.784\,mag in L\,1552. The median relative uncertainties of the resulting optical depth maps ($\tau_\mathrm{J}$) are 33.1\%, 12.1\%, and 28.4\%, respectively. 

In combination with the emission-derived column density map uncertainties (which were 19\% for CB\,68, 14\% in L\,429, and 16\% in L\,1552; see Section\,\ref{sec:emission_maps}), the median uncertainties in the comparison analysis were therefore 25.2\%, 16.2\%, and 23.6\%, respectively.

\subsection{Morphology} \label{sec:morphology}

Figure\,\ref{fig:map_compare} shows the extinction-derived optical depth maps ($\tau_\mathrm{J}$, left) and emission-derived optical depth maps at the fiducial wavelength 250\,$\micro$m ($\tau_{250}$, center-left) for the three cores. The center-right column shows the ratio of the two optical depth maps scaled by the theoretical opacity ratio for OH5a, $\left(\tau_\mathrm{J} / \tau_{250, \,\mathrm{OH5a}}\right)\left(\kappa_\mathrm{J} / \kappa_{250}\right)^{-1}_\mathrm{OH5a}$, and the right column describes the consistency of the scaled comparison as discussed in Section \ref{sec:map_consistency}. Filamentary extensions from the central regions are apparent in all three cores. As discussed in \citet{launhardt10}, there is a weak comet-like tail around CB\,68, apparent also in the RGB color image in Figure\,\ref{fig:rgb}. Similar to CB\,68, the cores L\,429 and L\,1552 have structures which extend away from their central cores, although they are more diffuse and broad. L\,429's tail primarily extends to the south and north-west of the core, whereas L\,1552's tail extends to the north.

\subsection{Optical depth map consistency} \label{sec:map_consistency}

\begin{figure*}[ht]
  \begin{center}
  \includegraphics[width=\linewidth]{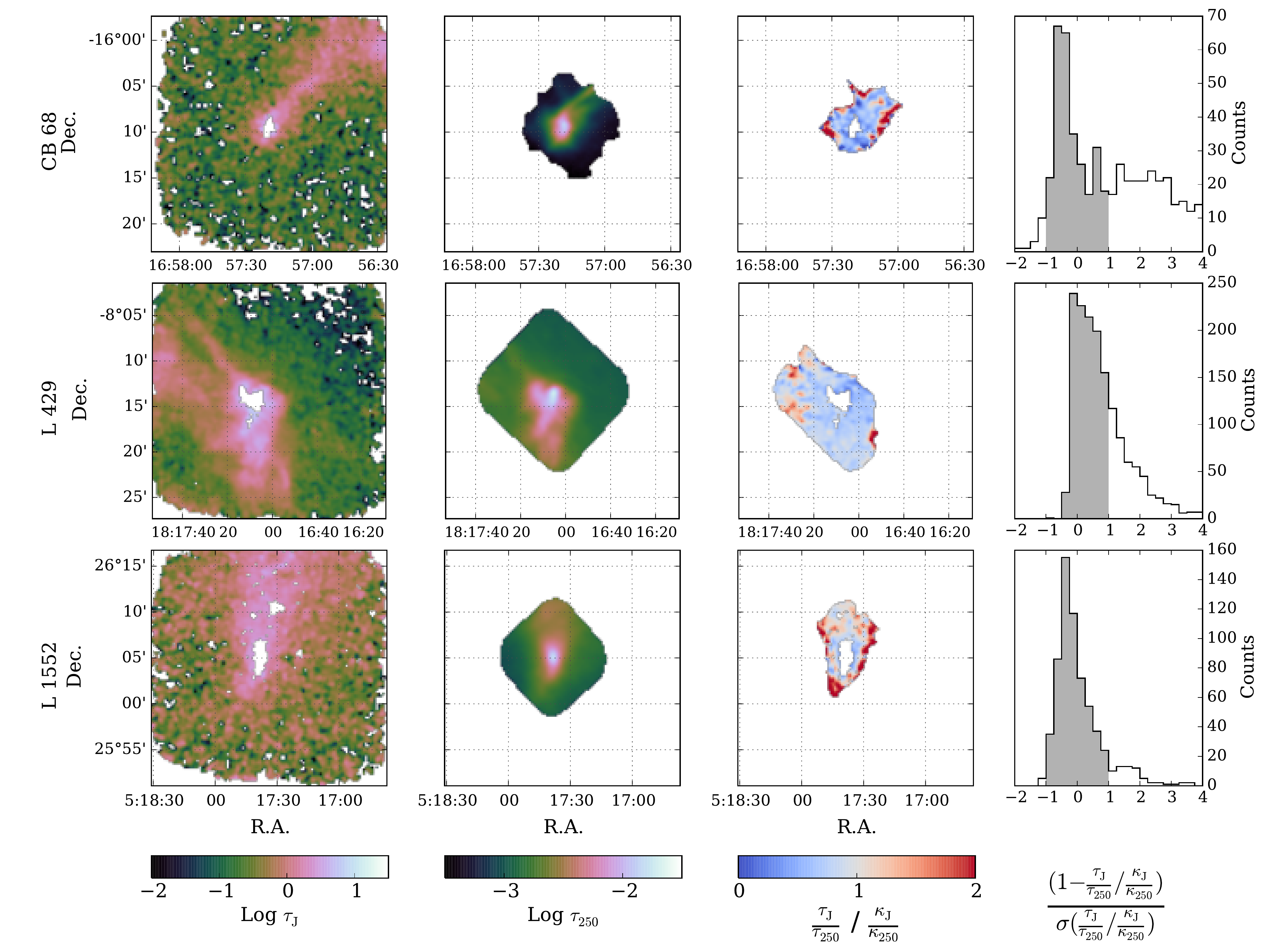}
  \end{center}
  \caption{Comparison between extinction-derived optical depth (far-left), and the emission-derived optical depth maps for OH5a at the fiducial wavelength 250\,$\micro$m (center-left) for CB\,68, L\,429, and L\,1552. The ratio of the optical depths $\tau_\mathrm{J} / \tau_{250}$ scaled by the theoretical opacity ratio from OH5a is shown on the center-right, where the agreement between the two maps is assessed through the difference of the ratio from unity over the formal uncertainty ($\xi$, see Equation \ref{eqn:consistency}). The $\xi$ distribution is shown on the far-right, grey region highlights where the $\xi$ distribution is consistent within 1\,$\sigma$, \textit{i.e.}, \%$\xi<$1$\sigma$. The statistics of the right-sided columns are listed in Table\,\ref{tab:consistency}, and on the vertical axis of the bottom row of Figure\,\ref{fig:tau_compare}. For reference, the optical depths are shown are those without background subtractions, although the regions below the background that is shown were masked in the comparison (center-right). } \label{fig:map_compare}
\end{figure*}

\begin{figure*}[ht]
  \begin{center}
  \includegraphics[width=\linewidth]{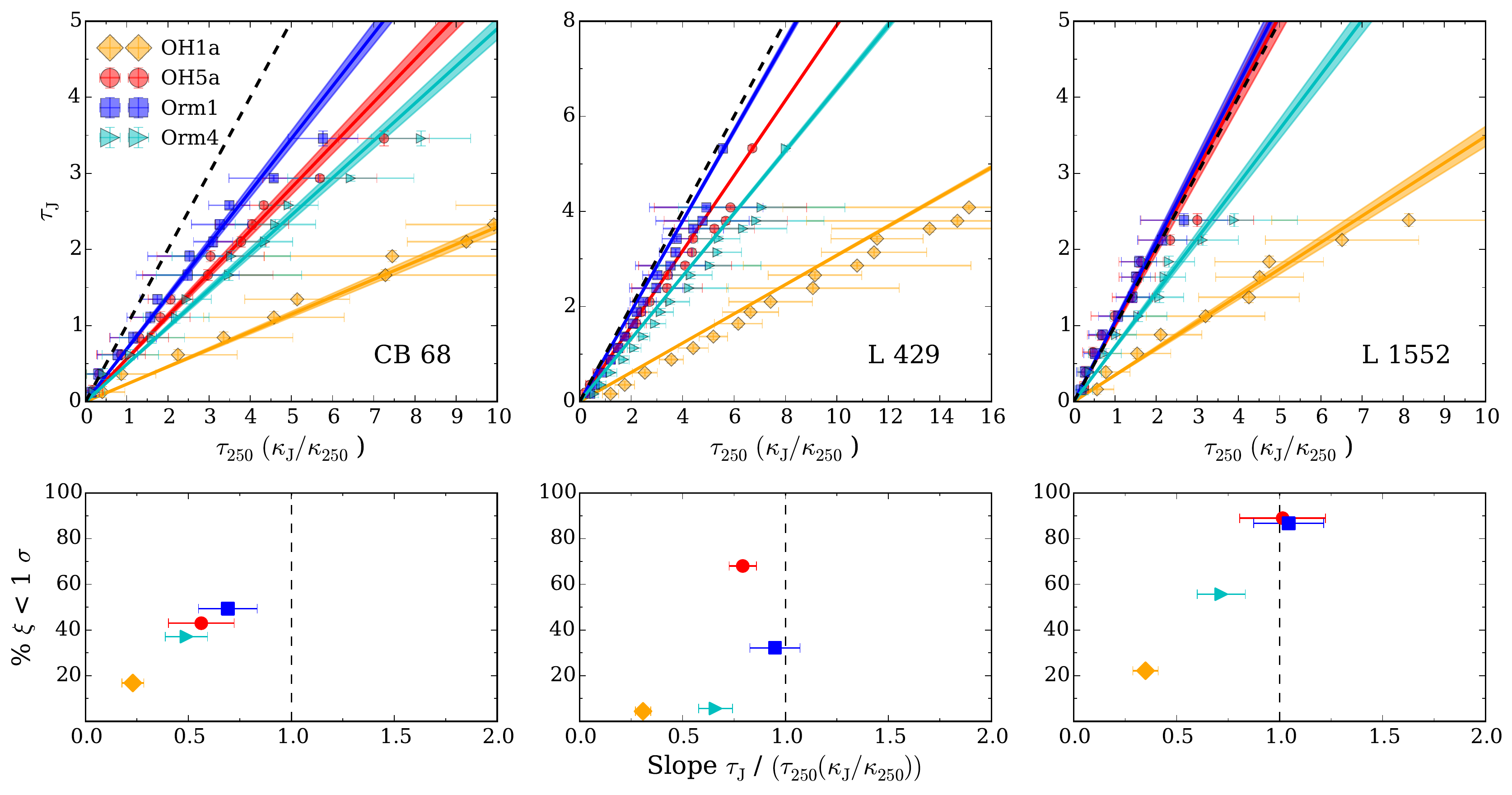}
  \end{center}
  \caption{Top row: Comparison between binned extinction- and emission-derived optical depths for each cloud and each model considered. The data points represent the median value of each bin, where the uncertainty is the standard deviation. Fitted slopes are shown with widths corresponding to the uncertainty on the fits. Bottom row: Summary of the extinction- and emission-derived optical depth comparison statistics, listed in Table\,\ref{tab:consistency}. On the horizontal axis, the uncertainty weighted slope of the scaled ratio is shown with errors corresponding to the standard deviation across the data points fit. The percent of the $\xi$ distribution within 1\,$\sigma$ is shown on the vertical axis. The model which produces an emission-derived optical depth map that is most comparable to the corresponding extinction-derived map will therefore be consistent with unity on the $x$-axis, and approach 100\% on the $y$-axis. } \label{fig:tau_compare}
\end{figure*}

\begin{table}
\centering
\begin{threeparttable}
\caption{Comparison of measured extinction- and emission-derived optical depth ratios, $\tau_\mathrm{J}/\tau_{250}$, with dust opacity ratios for several theoretical models. A visual representation of the data is shown in Figures\,\ref{fig:map_compare}\,and\,\ref{fig:tau_compare}. }
\label{tab:consistency}
\begin{tabular}{c|ccc}
\toprule
Model & slope $\tau_\mathrm{J} $/ $\left( \tau_{250} (\kappa_\mathrm{J}/\kappa_{250}) \right)$ & median $\xi$ & \% $<$ 1\,$\sigma$ \\
	  & (1) & (2) & (3) \\
\midrule
& \multicolumn{3}{c}{CB\,68} \\
\midrule  
  OH1a	& 0.252$\pm$0.006	& 5.4$\pm$9.3	& 16.8	\\
  OH5a	& 0.63$\pm$0.02	& 1.3$\pm$3.2	& 43.0	\\
  Orm1	& 0.75$\pm$0.02	& 0.4$\pm$2.5	& 49.3	\\
  Orm4	& 0.53$\pm$0.01	& 1.4$\pm$3.9	& 37.1	\\
  \midrule
& \multicolumn{3}{c}{L\,429} \\
\midrule
  OH1a	& 0.273$\pm$0.002	& 9.9$\pm$8.0	& 4.4		\\
  OH5a	& 0.743$\pm$0.006	& 0.6$\pm$1.4	& 68.0	\\
  Orm1	& 0.822$\pm$0.006	& 1.3$\pm$2.3	& 32.2	\\
  Orm4	& 0.576$\pm$0.004	& 3.0$\pm$3.4	& 5.6		\\
  \midrule
& \multicolumn{3}{c}{L\,1552} \\
\midrule
  OH1a	& 0.360$\pm$0.005	& 3.8$\pm$3.5	& 22.1	\\
  OH5a	& 1.07$\pm$0.02	& -0.2$\pm$0.8	& 89.0	\\
  Orm1	& 1.08$\pm$0.02	& -0.2$\pm$0.8	& 86.7	\\
  Orm4	& 0.74$\pm$0.01	& 0.7$\pm$1.3	& 55.7	\\
  \bottomrule
\end{tabular}
\begin{tablenotes}
      \footnotesize
      \item 1. Uncertainty-weighted slope of the ratio of the extinction- and emission-derived optical depth maps scaled by the expected slope. Errors correspond to the standard deviation of the fits determined from the covariance matrix.
      \item 2. Median of the comparison statistic (Equation \ref{eqn:consistency}), where the error is the standard deviation.
      \item 3. Percent of the $\xi$ distribution within 1\,$\sigma$-level.
    \end{tablenotes}
\end{threeparttable}
\end{table}

To describe the consistency between $\tau_{250}$ from thermal dust emission and $\tau_\mathrm{J}$ from dust extinction, we define the parameter $\xi$ as:
\begin{equation} \label{eqn:consistency}
\xi = \frac{1 - \left(\frac{\tau_\mathrm{J}}{\tau_{250}}/\frac{\kappa_\mathrm{J}}{\kappa_{250}}\right)}{\sigma\left( \frac{\tau_\mathrm{J}}{\tau_{250}}/\frac{\kappa_\mathrm{J}}{\kappa_{250}} \right) }
\end{equation}
\noindent where $\kappa_\mathrm{J}/\kappa_{250}$ comes from the adopted dust models (see Section\,\ref{sec:dust_models}) and $\sigma\left( \frac{\tau_\mathrm{J}}{\tau_{250}}/\frac{\kappa_\mathrm{J}}{\kappa_{250}} \right)$ is the formal uncertainty on the scaled ratio (hereafter, $\sigma$). This parameter represents a weighted scaled ratio for the optical depth maps for a given dust model.  A value of $\xi$=$0$ means the extinction maps and thermal dust emission maps have perfect agreement for a particular dust model.  

Figure\,\ref{fig:map_compare} shows histograms of $\xi$ for all three cores assuming OH5a dust. Table\,\ref{tab:consistency} lists the median $\xi$ value for all four dust models, as well as the percentage of $\xi$ values within 1\,$\sigma$ of $\xi$=$0$ (hereafter, \%$\xi<$1\,$\sigma$).  We use this percentage to assess the consistency of the individual dust models across all three sources.

The top row of Figure\,\ref{fig:tau_compare} shows comparisons of the extinction-derived optical depth with the scaled emission-derived optical depth for each cloud and each model on a pixel-by-pixel basis, where the data were binned with bin widths of 0.25\,$\tau_\mathrm{J}$. Only bins with more than five data points were considered, and the uncertainties on the data points correspond to one standard deviation in the values of each bin. A black dashed line of unity is shown for reference. A slope (zero intercept) was fit to $\tau_\mathrm{J} $/ $\left( \tau_{250} (\kappa_\mathrm{J}/\kappa_{250}) \right)$ for each model via orthogonal distance regression weighted by both uncertainty values, and is shown with a width corresponding to the error calculated from the covariance matrix of the fit. The fit was restricted to $\tau_\mathrm{J} > 1$ to reduce the bias of the low optical depth data.

The bottom row of Figure\,\ref{fig:tau_compare} shows that the consistency of the dust models differs substantially between the three cores. On the bottom horizontal axis, the slope of the binned and scaled ratio is shown with error bars corresponding to the standard deviation of the data. Although none of the models reach \%$\xi<$1\,$\sigma$ values above 50\% for CB\,68 (Orm1 is only slightly below this cutoff), OH5a is above 50\% for L\,429, and each of OH5a, Orm1, and Orm4 are above 50\% for L\,1552. Only Orm1 has a slope consistent with unity within one standard deviation for L\,429, however, and only Orm1 and OH5a have consistent slopes for L\,1552. Of particular note, OH1a is the least consistent of the models considered for each source.

Appendix\,\ref{app:other_models} introduces four additional models, two from \citet{ossenkopf94} and two from \citet{ormel11}, which describe properties similar to the present four models. In considering the full set of models, the results suggest that the optical properties of dust in the envelopes of the cores are best described by either silicate and bare graphite grains (like Orm1) or carbonaceous grains with some coagulation and either thin or no ice mantles (like OH5a). It is not unexpected that models with minimal ice formation are found to be most consistent in the envelopes of these cores. As noted in Section\,\ref{sec:dust_models}, the models assume densities of $\ttt{5}$\,cm$^{-3}$ whereas the material traced is probably closer to $\ttt{3}$-$\ttt{4}$\,cm$^{-3}$, where it will take longer for grains to coagulate.

Our results demonstrate the complex nature of determining the pre-stellar conditions of cores. Although we selected well-isolated and simply structured cores, models typically assumed to provide reasonable estimates of the dust properties of such objects do not consistently lead to correct total gas masses. Given that the models describe the center of the cores (whereas our comparison was restricted to the envelope), the consistency of some of the models was better than expected. The comparison highlights the limited capability of the SED fitting method, however \citep[as discussed by others, \textit{e.g.}, ][]{pagani15}. Alternatively, multiple temperature modified black body functions may be required to model the dust emission accurately. For example, differential coagulation and ice mantle growth between the cold dense core interiors and the thin warm outer envelopes may mean that single functions are not applicable \citep[as shown by studies of coreshine, \textit{e.g.},][]{steinacker14, lefevre14, lefevre16}.

\subsection{Improvement in uncertainties as a function of image depth} \label{sec:magdiff}

The analysis in this study could be improved through the reduction of the formal uncertainties in the optical depth maps, which could be achieved with more sophisticated SED modeling or deeper NIR observations. The uncertainties in the extinction values are a function of the intrinsic color scatter of the stellar population, photometric error, and the number of stars within a weighted Gaussian \citep{nicer}. Increasing the depth of the NIR observations would reduce the uncertainties twofold: it would reduce the photometric error, and increase the number of stars observed. Moreover, observations of a dedicated ``off'' position or larger maps that measure extinction-free locations would also help improve the NIR optical depth measurements. With smaller errors, an improved distinction could be made between the OH and Ormel models.

The potential improvements given 2 mag deeper observations were determined from artificially adding stars into our images with the appropriate number density, photometric errors, and color offset. The artificial photometry files were then used to produce extinction maps with the NICEST algorithm, and the errors were compared to the original maps. The simulated results had a $\sim$40\% improvement in the uncertainties. 

\subsection{Comparison with recent studies}

There has been a significant effort recently to constrain dust properties in star-forming regions. Although at resolutions too low to observe individual cores, observations with the \emph{Planck} telescope \citep[such as][]{pcxxiii, juvela15, ysard15} have yielded interesting results on the large-scale distribution of dust in many environments and their implications on the dust properties. In addition, many individual MCs have been studied. \citet{suutarinen13} studied the Corona Australis cloud and found a median extinction ratio of $A_{250}/A_\mathrm{J}$\,=\,$(1.4\pm0.2)\tttt{-3}$, or equivalently $\tau_\mathrm{J}/\tau_{250}=714\pm102$. Of the four dust models considered, this result is consistent with Orm1 only. Similarly, a study by \citet{juvela15} of 20 \emph{Herschel} fields found a median $\tau_\mathrm{J}/\tau_{250}=625\pm78$. With the \citet{bohlin78} column density relation and \citet{cardelli89} extinction curve, the \citet{planckxi14} result $\tau_{250}/\mathrm{N}_\mathrm{H}\sim0.55\tttt{-25}$\,cm$^2$\,H$^{-1}$ corresponds to $\tau_\mathrm{J}/\tau_{250}=2439$. Each of these studies, however, employ the relation $\kappa\propto\nu^2$ and thus are not entirely equivalent to our results. Regardless, the variance in values indicate the complex nature of this problem due to changes in dust properties at different densities within MCs and cores.

%%%%%%%%%%%%%%%%%%%%%%%%%%%%%%%%%%%%%%%%%%%%%%%%%%%%%%%%%%
%%%%%%%%%%%%%%%%%%%%%%%%%%%%%%%%%%%%%%%%%%%%%%%%%%%%%%%%%%

\section{CONCLUSIONS}
\label{sec:conclusions}

We studied the thermal dust emission and visual extinction of three nearby isolated cores to assess the validity of four dust opacity models used in the fit of the FIR dust continuum SEDs. The cores were carefully selected to be truly isolated, be previously well characterized, have exceptionally low FIR background continuum noise, and have a sufficiently high surface density of background stars from which deep NIR extinction maps can be derived.

The FIR maps were produced at five wavelength bands between 160\,$\micro$m and 500\,$\micro$m obtained with the \emph{Herschel} Space Observatory via the HSA. The individual SEDs in the map pixels were fit independently with a single-temperature modified blackbody curves using opacities from four (OH1a, OH5a, Orm1, and Orm4) selected dust property models. 

The NIR maps were obtained from the CFHT WIRCam in the J, H, and K bands. With the NICEST algorithm, a standard reddening law was used to measure the J-band extinction from the colors of the stars in a reddened region relative to an unextincted field. The extinction was then converted to optical depth via conversion constants.

We analyzed the extinction-derived maps with the four emission-derived optical depth maps to assess the validity of the dust models, for three cores. The maps were equivalently `background subtracted' prior to their comparison, where the background was measured at radii beyond 300\,$\arcsec$ of the central source. We obtained the following main results:

\begin{enumerate}

\item The slopes of the binned and scaled optical depth ratios, $\tau_\mathrm{J} $/ $\left( \tau_{250} (\kappa_\mathrm{J}/\kappa_{250}) \right)$, were consistent with unity within one standard deviation for none of the models for CB\,68, only Orm1 for L\,429, and both Orm1 and OH5a for L\,1552. 

\item The percentages of data within the error of the scaled ratio, \%$\xi<$1\,$\sigma$, were above 50\% for none of the models for CB\,68, only OH5a for L\,429, and each of Orm1, Orm4, and OH5a for L\,1552.

\item None of the models individually produces the most consistent optical depth maps for all three cores, suggesting the dust properties differ substantially between the sources. 

\item Unsurprisingly, the model describing ISM-like dust, OH1a, produces the least consistent optical depths for each of the cores.

\item As mentioned in Sections\,\ref{sec:herschel}\,and\,\ref{sec:wircam}, a simplistic approach was used to subtract the background emission and extinction, where the former was much more significant. Accordingly, the $\tau_{250}$ maps represent the lower limit of the comparison, and the upper limit of $\xi$ and $\tau_\mathrm{J} $/ $\left( \tau_{250} (\kappa_\mathrm{J}/\kappa_{250}) \right)$. In contrast, Appendix\,\ref{app:background} describes the lower limit of $\tau_\mathrm{J} $/ $\left( \tau_{250} (\kappa_\mathrm{J}/\kappa_{250}) \right)$, where no background subtraction was applied, and none of the models provide consistent optical depth maps. How the background emission is treated is therefore significant to the study of the dust properties.

% here2

\end{enumerate}

The results indicate that the dust in the cores is not well described by any individual dust property model considered, or cannot be fit by a single temperature modified black body function. Each dust model assumed here, however, describes dust at densities in the centers of cores ($\ttt{5}$\,cm$^{-5}$) yet our comparison is limited to envelopes ($\ttt{3}$\,cm$^{-3}$--$\ttt{4}$\,cm$^{-3}$) where longer timescales are required for dust coagulation and grain growth. The application of these dust models may be limited to the very center of the sources. 

As a next step, the distinction between the models could be improved with smaller errors, and provide greater confidence to the above conclusions. The presence of mid-IR scattering (`coreshine') in these cores may give an indication of the dust properties in their envelopes. Less-evolved dust grains could possibly be distinguished by such observations. 

\acknowledgments

\footnotesize{

We sincerely thank the referee for a very thorough and careful reading of the manuscript.

This work is supported in part by the NRC HIA and the University of Victoria, and is based on observations obtained with WIRCam, a joint project of Taiwan, Korea, Canada, France, and the CFHT which is operated by the National Research Council (NRC) of Canada, the Institut National des Sciences de l'Univers of the center National de la Recherche Scientifique of France, and the University of Hawaii. 

Additional observations were obtained from the \emph{Herschel} Space Observatory instruments PACS and SPIRE. 
The \emph{Herschel} spacecraft was designed, built, tested, and launched under a contract to ESA managed by the \emph{Herschel}/Planck Project team by an industrial consortium under the overall responsibility of the prime contractor Thales Alenia Space (Cannes), and including Astrium (Friedrichshafen) responsible for the payload module and for system testing at spacecraft level, Thales Alenia Space (Turin) responsible for the service module, and Astrium (Toulouse) responsible for the telescope, with in excess of a hundred subcontractors.

PACS has been developed by a consortium of institutes led by MPE (Germany) and including UVIE (Austria); KU Leuven, CSL, IMEC (Belgium); CEA, LAM (France); MPIA (Germany); INAF-IFSI/OAA/OAP/OAT, LENS, SISSA (Italy); IAC (Spain). This development has been supported by the funding agencies BMVIT (Austria), ESA-PRODEX (Belgium), CEA/CNES (France), DLR (Germany), ASI/INAF (Italy), and CICYT/MCYT (Spain). SPIRE has been developed by a consortium of institutes led by Cardiff University (UK) and including Univ. Lethbridge (Canada); NAOC (China); CEA, LAM (France); IFSI, Univ. Padua (Italy); IAC (Spain); Stockholm Observatory (Sweden); Imperial College London, RAL, UCL-MSSL, UKATC, Univ. Sussex (UK); and Caltech, JPL, NHSC, Univ. Colorado (USA). This development has been supported by national funding agencies: CSA (Canada); NAOC (China); CEA, CNES, CNRS (France); ASI (Italy); MCINN (Spain); SNSB (Sweden); STFC, UKSA (UK); and NASA (USA).

This work is based (in part) on data products produced at the TERAPIX data center located at the Institut d'Astrophysique de Paris. Data products from the Two Micron All Sky Survey are also used in this work, which is a joint project of the University of Massachusetts and the Infrared Processing and Analysis Center/California Institute of Technology, funded by the National Aeronautics and Space Administration and the National Science Foundation. 

\begin{sloppypar}
\software{I'iwi (www.cfht.hawaii.edu/Instruments/Imaging/WIRCam/IiwiVersion1Doc.html), \\
Source Extractor (v2.19.5; Bertin \& Arnouts 1996), \\
DIET (www.cfht.hawaii.edu/Instruments/Imaging/Megacam/dietmegacam.html)}
\end{sloppypar}

}

\bibliography{mcc_apj}

\normalsize

\appendix
\counterwithin{figure}{section}

\section{Significance of the background subtraction}\label{app:background}

\begin{figure*}[ht]
  \begin{center}
  \includegraphics[width=\linewidth]{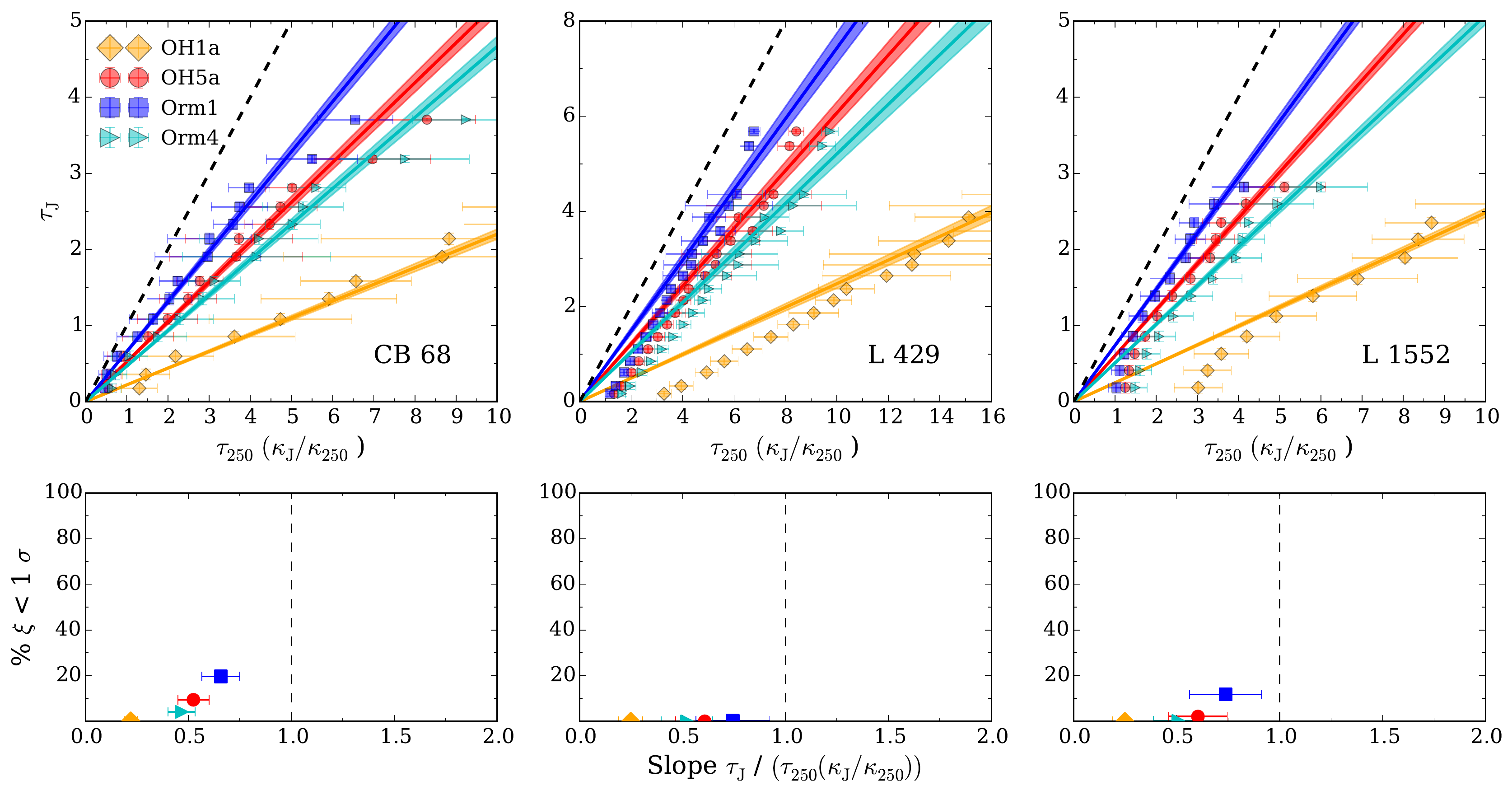}
  \end{center}
  \caption{Top row: Comparison between binned extinction- and emission-derived optical depths for each cloud and each model considered, where no background subtraction has been performed. The data points represent the median value of each bin, where the uncertainty is the standard deviation. Fitted slopes are shown with widths corresponding to the uncertainty on the fits. Bottom row: Summary of the extinction- and emission-derived optical depth comparison statistics. On the horizontal axis, the uncertainty weighted slope of the scaled ratio is shown with errors corresponding to the standard deviation across the data points fit. The percent of the $\xi$ distribution within 1\,$\sigma$ is shown on the vertical axis. The model which produces an emission-derived optical depth map that is most comparable to the corresponding extinction-derived map will therefore be consistent with unity on the $x$-axis, and approach 100\% on the $y$-axis. } \label{fig:tau_compare_app}
\end{figure*}

As described in Sections\,\ref{sec:herschel} and \ref{sec:wircam}, an estimate of the background emission was determined from the median flux levels beyond 300\,$\arcsec$ radii of the central core for each of the maps. The background extinction levels were similarly determined from the extinction maps, yet were found to be negligible relative to the emission background. These backgrounds were subtracted prior to the comparison of the optical depths, where the above results represent upper limits for the $\tau_\mathrm{J}$ to $\tau_{250}$ comparisons. To emphasize the necessity of performing this background subtraction, Figure\,\ref{fig:tau_compare_app} shows the equivalent comparison as Figure\,\ref{fig:tau_compare} where the background was not subtracted, \textit{i.e.}, the lower limit of $\tau_\mathrm{J} $/ $\left( \tau_{250} (\kappa_\mathrm{J}/\kappa_{250}) \right)$. Neglecting to correct for background emission (which dominates background extinction) results in slopes significantly ($\sim$34--53\%) lower for each of the models. In this extreme case, none of the dust models produce consistent optical depths by any of our measures.

\section{Additional dust models} \label{app:other_models}

Additional dust models that we considered were two from \citet[][OH2a and OH8a]{ossenkopf94} and two from \citet[][Orm2 and Orm3]{ormel11}. Each model describes a mix of the dust properties of the models in the main text. The 1.25\,$\micro$m and 250\,$\micro$m opacities for these models are listed in Table \ref{tab:kappas_add}. The same background subtraction procedure as described in Section\,\ref{sec:herschel} was applied to the emission maps.

The model OH2a describes silicate-based carbonaceous dust grains without ice mantles, and gas densities of $\ttt{5}$\,cm$^3$. OH8a, on the other hand, describes the same grains but with thick ice mantles coagulated over $\ttt{5}$\,years. For both these models, as with the previous OH models, the opacities were converted to extinction opacities (refer to Section\,\ref{sec:dust_models}). The Orm models were derived for mixed silicate and graphite grains of fraction 2:1 and ice mantle thicknesses coagulated for $\ttt{5}$\,years. Specifically, Orm2 has ice-silicate and ice-graphite grains, and Orm3 has ice-silicate and bare graphite grains. Orm3 differs from Orm4 in that the mixing is between grains of the same material, rather than on the level of individual aggregates.

\begin{figure*}[ht]
  \begin{center}
  \includegraphics[width=\linewidth]{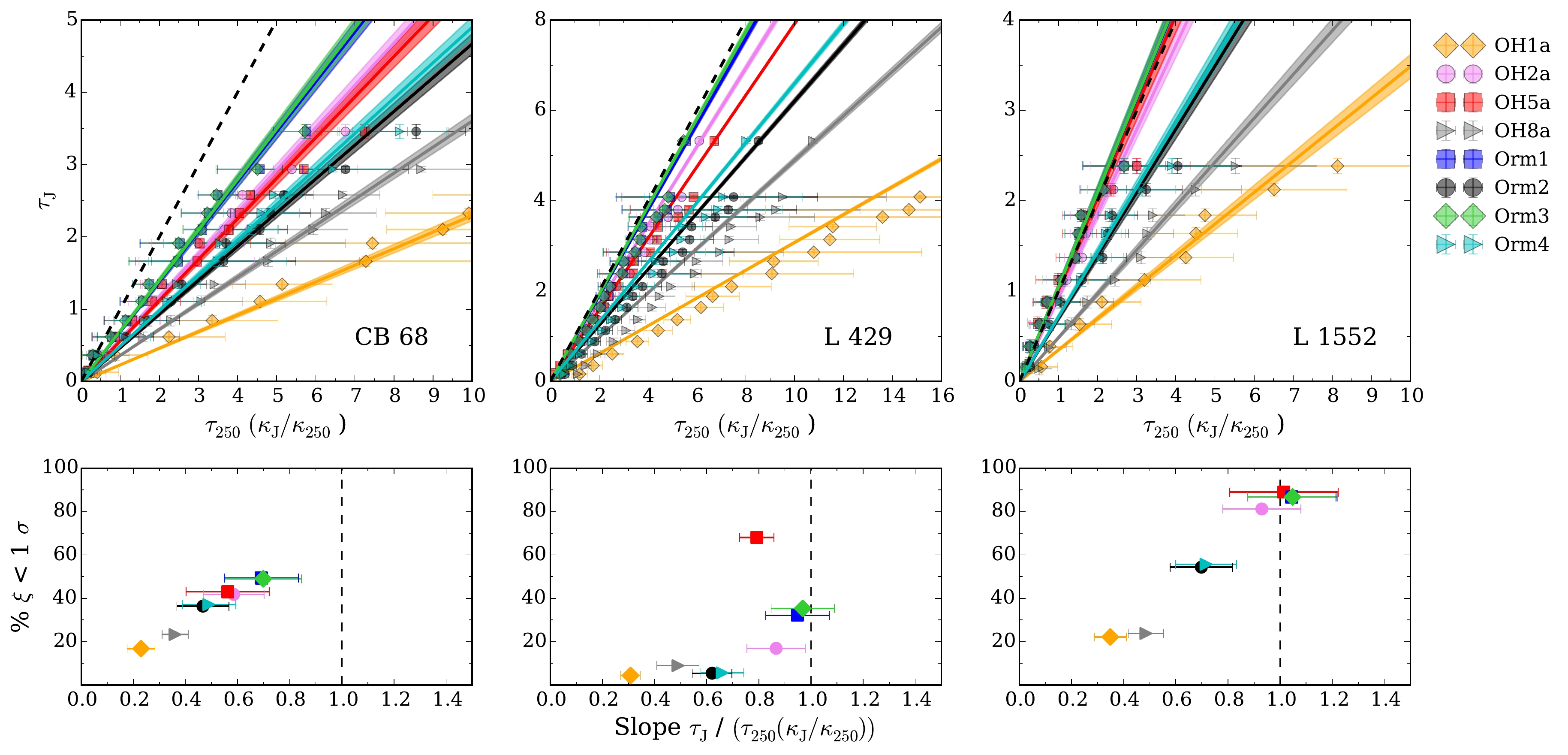}
  \end{center}
  \caption{Top row: Comparison between binned extinction- and emission-derived optical depths for each cloud and each model considered. The data points represent the median value of each bin, where the uncertainty is the standard deviation. Fitted slopes are shown with widths corresponding to the uncertainty on the fits. Bottom row: Summary of the extinction- and emission-derived optical depth comparison statistics. On the horizontal axis, the uncertainty weighted slope of the scaled ratio is shown with errors corresponding to the standard deviation across the data points fit. The percent of the $\xi$ distribution within 1\,$\sigma$ is shown on the vertical axis. The model which produces an emission-derived optical depth map that is most comparable to the corresponding extinction-derived map will therefore be consistent with unity on the $x$-axis, and approach 100\% on the $y$-axis. } \label{fig:tau_compare_all}
\end{figure*}

In repeating the analysis described in Section\,\ref{sec:map_consistency}, Figure\,\ref{fig:tau_compare_all} shows the result of the comparison between extinction- and emission-derived opacities. The consistency of the maps for Orm2 is comparable to that of Orm4, and Orm3 is comparable to Orm1. Although the OH2a maps are nearly as consistent as for OH5a for CB\,68 and L\,1552, the consistency is $\sim$50\%$\xi<$1\,$\sigma$ and lower for L\,429 despite the slope being slightly nearer unity. The OH8a maps are relatively inconsistent for all three sources, although not as much as the OH1a maps.

The results from the full set of models suggest that the optical properties of the dust in the envelopes of the three sources are most closely described by either silicate and bare graphite grains (Orm1 or Orm3) or carbonaceous grains with some coagulation and either thin or no ice mantles (OH5a or OH2a). On one level, this makes sense since the models assume densities of $\ttt{5}$\,cm$^{-3}$ whereas the material traced is probably closer to $\ttt{3}$-$\ttt{4}$\,cm$^{-3}$ in density.  It will take longer for grains to coagulate at these lower densities.

\begin{table}
\centering
\begin{threeparttable}
\caption{Dust optical depth values for the additional models considered. Values are interpolated }
\label{tab:kappas_add}
\begin{tabular}{ccc}
\toprule
  Model	& $\kappa_\mathrm{J}$ [$\tttt{4}$ cm$^2$g$^{-1}$]	& $\kappa_{250}$ [cm$^2$g$^{-1}$] 	\\
 \midrule
OH2a	& 1.817	& 10.78	 \\
OH8a	& 2.162	& 18.81	 \\
Orm2	& 1.264	& 12.52	 \\
Orm3	& 1.786	& 20.78	 \\
\bottomrule
\end{tabular}
\end{threeparttable}
\end{table}

\end{document}